\RequirePackage{fix-cm}
\documentclass[smallextended]{svjour3}       
\smartqed  
\usepackage{graphicx}
\usepackage{mathptmx}      
\usepackage{geometry}                
\usepackage{graphicx}
\usepackage{epstopdf}
\usepackage{xargs}
\usepackage{ifthen}
\usepackage{extpfeil}
\usepackage{tikz}
\usepackage{dsfont}
\usepackage{pgf}
\usepackage{stmaryrd}
\usepackage[verbose]{wrapfig}
\usepackage{xcolor}
\usepackage{pbox}
\usepackage{url}
\usetikzlibrary{arrows}
\usepackage{amssymb}
\usepackage{amsmath}
\usepackage{soul}
\usepackage{verbatim}
\usepackage{hyperref}
 \usepackage[percent]{overpic}
\usepackage{algorithm}
\usepackage[noend]{algpseudocode}
\usepackage[T1]{fontenc}   
\usepackage{cite} 

\begin{document}

\title{Building an iterative heuristic solver for a quantum annealer \thanks{This research was supported by 1QB Information Technologies (1QBit) and Mitacs.
}
}


\author{Gili Rosenberg         	\and
        Mohammad Vazifeh		\and
        Brad Woods			\and
        Eldad Haber 
}


\institute{{G. Rosenberg \and M. Vazifeh \and B. Woods} \at
	      1QB Information Technologies (1QBit), Suite 900, 609 W Hastings Street, Vancouver, British Columbia V6B 4W4, Canada \\
              Tel.: +1-646-820-8865\\
              \email{gili.rosenberg@1qbit.com}           
           \and
           M. Vazifeh \at
              \email{mmvazifeh@gmail.com}           
           \and
	  B. Woods \at
	     \email{brad.woods@1qbit.com}
	  \and
	  E. Haber \at
	  Department of Mathematics and Earth and Ocean Science, University of British Columbia, Vancouver, British Columbia V6T 1Z2, Canada \\
	    \email{haber@math.ubc.ca}
}

\date{Received: date / Accepted: date}

\maketitle

\begin{abstract}
A quantum annealer heuristically minimizes quadratic unconstrained binary optimization (QUBO) problems, but is limited by the physical hardware in the size and density of the problems it can handle. We have developed a meta-heuristic solver that utilizes D-Wave Systems' quantum annealer (or any other QUBO problem optimizer) to solve larger or denser problems, by iteratively solving subproblems, while keeping the rest of the variables fixed. We present our algorithm, several variants, and the results for the optimization of standard QUBO problem instances from OR-Library of sizes 500 and 2500 as well as the Palubeckis instances of sizes 3000 to 7000. For practical use of the solver, we show the dependence of the time to best solution on the desired gap to the best known solution. In addition, we study the dependence of the gap and the time to best solution on the size of the problems solved by the underlying optimizer.
\end{abstract}


\section{Introduction}


\subsection{The problem and previous work}

The quadratic unconstrained binary optimization (QUBO) problem is defined by
 \begin{eqnarray*} \min & x^TQx \\ s.t. & x \in S, \end{eqnarray*} 
where, without loss of generality, $Q\in \mathbb{R}^{N\times N}$ and $S$ represents the binary discrete set $\{ 0,1\}^N$.

Many NP-hard combinatorial optimization problems arise naturally or can easily be reformulated as QUBO problems, such as the quadratic assignment problem, the maximum cut problem, the maximum clique problem, the set packing problem, and the graph colouring problem (see, for instance, Boros and Pr{\'e}kopa~\cite{Boros:1989}, Boros and Hammer~\cite{Boros:2002}, Bourjolly et al.~\cite{Bourjolly:1994}, Du and Pardalos~\cite{Du:1999}, Pardalos and Rodgers~\cite{Pardalos:1990,Pardalos:1992}, Pardalos and Xue~\cite{Pardalos:1994}, and Kochenberger et al.~\cite{Kochenberger:2004}).

Numerous interesting applications expressed naturally in the form of a QUBO problem have appeared in the literature.  Barahona et al.~\cite{Barahona:1988, deSimone:1995} formulate and solve the problem of finding exact ground states of spin glasses with magnetic fields.  Alidaee et al.~\cite{Alidaee:1994} study the problem of scheduling $n$ jobs non-preemptively on two parallel, identical processors to minimize weighted mean flow time as a QUBO problem.  Bomze et al.~\cite{Bomze:1999} give a comprehensive discussion of the maximum clique (MC) problem.  Included is the QUBO problem representation of the MC problem and a variety of applications from different domains.  The QUBO problem has been used in the prediction of epileptic seizures~\cite{Iasemidis:2001}. Alidaee et al.~\cite{Alidaee:2005} discuss the number partitioning problem as a QUBO problem.

To solve a QUBO problem, a number of exact methods have been developed~\cite{Gulati:1984,Carter:1984,Williams:1985,Barahona:1989,Pardalos:1990,Pardalos:1990b, Billionnet:1994,Palubeckis:1995,Helmberg:1998,Hansen:2000,Huang:2006,Pardalos:2006,Pan:2008,Gueye:2009,Pham-Dinh:2010,Mauri:2011,Mauri:2012a,Mauri:2012b,Li:2012}.  However, due to the computational complexity of the problems, these approaches have only been able to solve small-sized instances.  To obtain near-optimal solutions, several heuristic and metaheuristic approaches such as tabu search~\cite{Beasley:98,Glover:1998,Glover:1999,Palubeckis:2006,Glover:2010,Lu:2010,Shylo:2011,Lu:2011,Wang:2013}, simulated annealing~\cite{Alkhamis:1998,Katayama:2001}, evolutionary algorithms~\cite{Merz:1999,Katayama:2000,Lodi:1999,Lu:2010,Cai:2011,Wang:2012a}, scatter search~\cite{Amini:1999}, GRASP~\cite{Palubeckis:2002,Wang:2013}, and iterated local search algorithms~\cite{Boros:2007} have been proposed.  Additionally, preprocessing techniques have been developed to simplify the solving of QUBO problems~\cite{Boros:2006}.

The majority of existing heuristic algorithms search relatively small neighbourhoods that can be searched efficiently.  In this paper, we explore the possibility of utilizing a quantum annealer (see Section~\ref{QuantumAnnealer}), which is able to search a portion of the $k$-flip neighbourhood of size $2^k$ in a single operation.  We develop a hybrid algorithm that incorporates a quantum annealer into a classical algorithm and give experimental results that show that this approach could be competitive in the near future.

While working on this research we noted research on the possibility of solving larger problems by decomposing them into smaller problems and solving them on the D-Wave machine \cite{bian2014discrete}. Although the idea of ``large-neighbourhood local search'' is mentioned in brief in that research, which bears resemblance to the basic idea we cover in this paper, it is not developed in detail and no results are reported. 

After the publication of our paper on the arXiv, we became aware of a paper which explores a similar idea \cite{zintchenko2015local}. The authors of that work explored the applicability of solution via local updates to two- and three-dimensional Ising spin glasses, as well as Chimera-structured problems. Since those problems contain only very local couplings, even a simple approach showed success. Our contribution focuses on devising an algorithm that is effective for a more general class of problems, sparse or dense, which in general contain short- and long-range couplings.

\subsection{The quantum annealer}
\label{QuantumAnnealer}

D-Wave Systems has developed a scalable quantum annealer that solves QUBO problems\footnote{The problem that the machine actually solves is known as an
Ising model in the physics community, and can easily be transformed into the form $x^T Q x$,
which is more common in the scientific computing community.}, 
albeit with some limitations. 

It has been argued that quantum annealing has an advantage over simulated annealing due to quantum tunnelling, which allows an optimizer to pass
through barriers instead of going over them. This might provide a speedup for some problem classes \cite{kadowaki1998quantum, finnila1994quantum, ray1989sherrington}. Indeed, simulated quantum 
annealing \cite{santoro2002theory, martovnak2002quantum} shows a speedup for some problem
classes \cite{santoro2002theory}, but not for some others \cite{battaglia2005optimization}.  

There is strong evidence that the D-Wave machine is quantum  \cite{lanting2014entanglement, boixo2014computational}, but it is yet to be established whether it provides a quantum speedup over classical computers. Recently, there has been significant 
interest in benchmarking the D-Wave machines \cite{mcgeoch2013experimental, katzgraber2014glassy, boixo2014evidence, hen2015probing, king2015performance}, and there is ongoing debate on how to define quantum speedup, and on which problems a quantum annealer could be expected to demonstrate it \cite{ronnow2014defining, martin2015unraveling, katzgraber2015seeking}. We expect results in the near future to shed light on this important point.

The connectivity of the D-Wave processor's qubits is currently given by the Chimera graph \cite{bunyk2014architectural}.
Due to the sparsity of this hardware graph, it is often necessary to identify multiple physical qubits with one logical qubit 
in order to increase the connectivity. In particular, it is known that the largest complete graph that can be embedded on a Chimera graph of size $n$ is of size $\sqrt{2n}+1$ (for instance, for the current-generation 512-qubit chip\footnote{ The chip is current as of mid-July 2015.}, the largest complete graph is of size 33) \cite{choi2008minor, choi2011minor}. The parameter misspecification error rate for the current-generation chip is 3--5$\%$, whereas the next
generation of chip, scheduled for release in the second half of 2015, will have
1152 active qubits and the error rate reduced by an expected $33\%$ \cite{williams2014}. 

The remainder of this paper is organized as follows. In Section~\ref{gqs} we describe the algorithm. In Section~\ref{benchmarking} we explain the benchmarking methodology and present computational results. In Sections~\ref{discussion} and~\ref{conclusions} we discuss the results, the conclusions, and future work.


\section{The General QUBO Problem Solver (GQS)}
\label{gqs}


\subsection{The basic algorithm}

\label{theidea}
We assume that we have an oracle that optimizes QUBO problems of size $k$, but our problem has $N \gg k$ variables. How can we use the oracle to solve this larger problem? Our idea, which we refer to as ``convergence'' henceforth, is to iteratively choose $k$ variables out of the $N$ variables, and optimize over those variables while keeping the other $N-k$ variables fixed, until the value of the objective function does not change for some number of steps. This basic idea can be improved upon significantly, by choosing intelligently which variables to update, for example, or using tabu search to disallow updating of the same variables too often, and so on (various improvements are described in Section~\ref{improvedalgorithm}). One can write a simple algorithm that uses this idea to find a local minimum for a problem of size $N$ (see Algorithm~\ref{BasicAlgorithm}).

\begin{algorithm}
\vspace*{2 mm}
\caption{Basic Algorithm}\label{BasicAlgorithm}
\begin{algorithmic}[1]
\Procedure{BasicAlgorithm}{}

\State $x \gets$ a random N-bit configuration, $x \in \{ 0,1\}^N$

\While {not converged}

\State Choose $k$ variables at random

\State $x \gets$ optimize over $k$ variables, given the fixed $N - k$ variables (using the underlying optimizer)

\EndWhile

\State{\Return $x$}

\vspace*{2 mm}

\EndProcedure
\end{algorithmic}
\end{algorithm}

Each call to the underlying optimizer and the subsequent update will either decrease the objective function's value or leave it constant, but is guaranteed not to increase it. This does not mean that the algorithm does not hill climb. It does, in fact, do so, but the hill climbing capability is contained within the optimizer of size $k$. This algorithm successfully finds a local minimum; however, we are searching for the global minimum, and hence we must improve the algorithm to improve our chances of success. 


\subsection{The improved algorithm}
\label{improvedalgorithm}

We here describe an improved version of Algorithm~\ref{BasicAlgorithm}, 
which we refer to as the General QUBO Problem Solver (GQS). The pseudo-code is presented in Algorithm~\ref{ImprovedAlgorithm}, and each of the components is described in greater detail below:
\begin{enumerate}
\item Better initial guesses (Section~\ref{InitialGuesses})
\item Different underlying optimizers (Section~\ref{DifferentMiniSolvers})
\item Escape strategies (Section~\ref{Escapes})
\item Choice of variables (Section~\ref{VariableChoice})
\item Tabu search for $k$-opt (Section~\ref{tabukopt})
\item 1-opt tabu search improvement (Section~\ref{Tabu1optImprovement})
\item Stopping criteria and convergence criterion (Section~\ref{StoppingCriteria})
\end{enumerate}

Each of these components is added in a modular way, such that many variations on the algorithm can be accommodated within this same framework. The reasoning behind this is that we expect that for different classes of problems the best-performing variant of this algorithm will be different. In addition, the mode of use is important: the parameters would be different if what we were looking for is a quick solution as opposed to a high rate of success. 

We note that the reason to keep track of the one-flip gains\footnote{A one-flip gain is the change in the objective function's value given a flip of a single bit. We refer to the collection of all possible single flips and their corresponding change in the objection function's value as the ``one-flip gains''.} is that once they are initialized, they can be updated very efficiently \cite{Merz2004, glover2010efficient}, which allows one to both quickly update the objective function's value after each $k$-variable update, and to use the one-flip gains to choose the variables to be updated (see Section~\ref{gainsbased}).

\begin{algorithm}
\caption{Improved Algorithm}\label{ImprovedAlgorithm}
\begin{algorithmic}[1]
\vspace*{2 mm}
\Procedure{ImprovedAlgorithm}{}

\State $V, x \gets$ InitialGuess()

\State $V_{\tt{min}}, x_{\tt{min}} \gets V, x$

\State $C \gets \{1, ..., N\}$

\State $T \gets \{\}$

\State gains $\gets$ Calculate change in objective function value for flipping each bit
\While {not Done()}

\State ChosenVariables $\gets$ VariableChoice($C, k$)

\State $x \gets$ Optimize over the $k$ ChosenVariables, given the fixed $N-k$ variables (using the underlying optimizer)

\State Set ChosenVariables in $x$ to the values given by the previous step

\State Update one-flip gains and $V$

\State $C, T \gets$ TabuUpdate()

\If {$V<V_{\tt{min}}$}
\State $V_{\tt{min}} \gets V$
\State $x_{\tt{min}} \gets x$
\EndIf

\If{Converged()} 
\State $x \gets$ Escape()
\State Update one-flip gains and $V$
\EndIf

\EndWhile

\State \Return $V_{\tt{min}}, x_{\tt{min}}$

\vspace*{2 mm}

\EndProcedure
\end{algorithmic}
\end{algorithm}


\subsection{Better initial guesses}
\label{InitialGuesses}

We use the greedy heuristic given by Merz and Katayama~\cite{Merz2004} to find good (i.e. with low value and diverse) starting configurations quickly. First, all variables are initialized to $0.5$, then repeatedly flipped to either $0$ or $1$ depending on which update provides the largest one-flip gain in the objective function's value, and then updating the one-flip gains and repeating, while updating each bit only once. The bits are initialized to $0.5$ despite being binary (that is, having values of $0$ or $1$) since the idea is not to bias the configurations initially such that any of the bits will be more likely to be $0$ or $1$ (by the end of the process all of the bits are set to either $0$ or $1$). 

When initializing the reference set, we experimented with using a deterministic version of this greedy heuristic for the first element in the reference set (see Algorithm~\ref{DeterministicGreedy}). In order to obtain a diverse reference set, for the remaining elements we used a randomized version, where a random bit is flipped randomly, and at each stage the best flip to 0 or 1 is chosen with a probability that depends on the best one-flip gains. Although this results in a diverse set of initial solutions, we found that starting with random configurations gave more diverse initial sets (for example, based on the average Hamming distance).

We note that the randomized version by Merz and Katayama \cite{Merz2004} only works when the highest 1-flip gains for both the best 0-flip and the best 1-flip are positive (for example, if they are both negative, $p$ can be negative). In addition, in that version the 1-flip gains are not updated after the initial random bit is flipped. We also note that our version is for minimization, not maximization. We present our corrected version of the pseudo-code; see Algorithm~\ref{RandomizedGreedy}. 

\makeatletter
\def\BState{\State\hskip-\ALG@thistlm}
\makeatother

\begin{algorithm}
\caption{Randomized Greedy Initial Guess}\label{RandomizedGreedy}
\begin{algorithmic}[1]
\vspace*{2 mm}
\Procedure{RandomizedGreedy}{}

	\State $C \gets \{1,...,N\}$
	\For {$i \in \{1,...,N\}$} $x_i = \frac12$
	\EndFor

\State Calculate one-flip gains $g_i$ for all $i \in \{1,...,N\}$
\State Select $k \in \{1,...,N\}$  and $l \in \{0, 1\}$ randomly and set $x_k \gets l$
\State $C \gets C - \{k\}$

\Repeat 
\State Update one-flip gains $g_i$ for all $i \in C$
\State Find $k_0$ with $g_k^0 =$ min$_{i \in C} \, g_i^0$ and $k_1$ with $g_k^1 =$ min$_{i \in C} \, g_i^1$

	\If {$g_{k_0}^0 < 0$ and $g_{k_1}^1 \leq 0$}
				$p \gets \frac{g_{k_0}^0}{g_{k_0}^0 + g_{k_1}^1}$
	\ElsIf {$g_{k_0}^0 \geq 0$ and $g_{k_1}^1 > 0$}
				$p \gets \frac{g_{k_1}^1}{g_{k_0}^0 + g_{k_1}^1}$
	\ElsIf {$g_{k_0}^0 > 0$ and $g_{k_1}^1 < 0$}
				$p \gets 0$
	\ElsIf {$g_{k_0}^0 < 0$ and $g_{k_1}^1 > 0$}
				$p \gets 1$
	\ElsIf {$g_{k_0}^0 = 0$ and $g_{k_1}^1 = 0$}
				$p \gets \frac12$
	
	\EndIf
		
	\If {random($0,1$) < $p$}
		\State{$x_{k_0} \gets 0$; $C \gets\ C - \{k_0\}$}
	\Else
		\State{$x_{k_1} \gets 1$; $C \gets C - \{k_1\}$}
	\EndIf
\Until {$C = \{\}$}

\State{\Return $x$}
\vspace*{2 mm}

\EndProcedure
\end{algorithmic}
\end{algorithm}

\begin{algorithm}
\caption{Deterministic Greedy Initial Guess}\label{DeterministicGreedy}
\begin{algorithmic}[1]
\vspace*{2 mm}
\Procedure{DeterministicGreedy}{}

	\State $C \gets \{1,...,N\}$
	\For {$i \in \{1,...,N\}$} $x_i = \frac12$
	\EndFor

\State Calculate one-flip gains $g_i$ for all $i \in \{1,...,N\}$

\Repeat 

\State Find $k_0$ with $g_k^0 =$ min$_{i \in C} \, g_i^0$ and $k_1$ with $g_k^1 =$ min$_{i \in C} \, g_i^1$

	\If {$g_{k_0}^0 < g_{k_1}^1$}
		\State{$x_{k_0} \gets 0$; $C \gets\ C - \{k_0\}$}		
	\Else
		\State{$x_{k_1} \gets 1$; $C \gets\ C - \{k_1\}$}		
	\EndIf
	
	\State Update one-flip gains $g_i$ for all $i \in C$;

\Until {$C = \{\}$}

\State{\Return $x$}

\EndProcedure
\end{algorithmic}
\end{algorithm}


\subsection{Different underlying optimizers}
\label{DifferentMiniSolvers}

The proposed GQS algorithm lends itself to being solved using any appropriate optimizer; for example, an exhaustive search, a 1-opt tabu search, Gurobi Optimizer, CPLEX, or a quantum annealer. The algorithm has been successfully verified using the D-Wave II quantum annealer. However, since annealing technology is at an early stage, limitations such as qubit connectivity, number of qubits, and noise and error levels restrict the size of problem that can be solved. For this reason, for our benchmarking we used a 1-opt tabu search, with a tabu tenure of 15--20 and convergence length of $10k$, where $k$ is the size of the underlying optimizer. We have verified that the GQS algorithm works when using the quantum annealer as the underlying optimizer. 


\subsection{Escape strategies}
\label{Escapes}

Once the local search phase converges, which consists of multiple calls to the underlying optimizer, the GQS attempts to ``escape'' by flipping some number of bits, and then performing another local search phase. If the solver converges to a different configuration, then the escape has succeeded in diversifying the search space of the algorithm. The most trivial example of an escape is to flip a fixed number of bits randomly. We describe some more sophisticated ways of escaping below. 


\subsubsection{Path relinking}
\label{PathRelinking}

Path relinking refers to maintaining a collection of the best solutions to which the solver converged, also referred to as a ``reference set'', and then escaping to solutions found by fusing two elite solutions that have not been fused before. This method is used in Wang et al.~\cite{wang2012path}, utilizing a tabu 1-opt local search.

Algorithm~\ref{PathRelinkingEscape}, ``Path Relinking Escape'', is composed of two phases. In the first, we maintain the reference set given a new converged solution. If the converged solution already appears in the reference set, we discard it. Otherwise, if the reference set is not full we add the converged solution, and if the reference set is full we only add the solution if it is better than the worst solution in the reference set, which we then discard. In the second phase, we perform the escape. If the reference set is full and unfused pairs remain with a sufficiently large Hamming distance, we choose an unfused pair randomly, and update the current solution to the fused ``child''. If the reference set is not yet full, we escape to a random configuration. Once we run out of unfused pairs, we empty the reference set except for the best solution thus far found, and restart the procedure. 

There are different ways to fuse the parents. We choose to ignore pairs of parents that have a Hamming distance $d$ less than a parent distance threshold (typically a small number such as 5). For parents with a Hamming distance greater than that, we identify the bits that are equal and the ones that are different in both parents. First, we set the equal bits in the child to be equal to those in the parents. Then we set the different bits in the child randomly, such that the resulting child's Hamming distance to each of the parents is at least a user-defined fraction of the parent-parent Hamming distance (for example, 0.3). 

\begin{algorithm}
\caption{Path Relinking Escape}\label{PathRelinkingEscape}
\begin{algorithmic}[1]
\vspace*{2 mm}
\Procedure{PathRelinkingEscape}{}

\If {$x \notin$ \detokenize{ref_set}}
	\If {\detokenize{ref_set} not full}
		\State{Add $x$ to \detokenize{ref_set}}
		\State {Update \detokenize{unfused_pairs}}
	\Else
		\If {$V$ is better than worst in \detokenize{ref_set}}
			\State{Pop worst from \detokenize{ref_set}}
			\State{Add $x$ to \detokenize{ref_set}}
			\State {Update \detokenize{unfused_pairs}}
		\EndIf
	\EndIf
\EndIf

\If {\detokenize{ref_set} is full}
	\If{\detokenize{unfused_pairs} is not empty}
		\State{$x \gets$ Fuse random pair from \detokenize{unfused_pairs}}
	\Else
		\State{Empty \detokenize{ref_set} except best}
		\State{Empty \detokenize{unfused_pairs}}
		\State{$x \gets$ RandomizedGreedy()}
	\EndIf
\Else { $x \gets$ RandomizedGreedy()}
\EndIf

\State {\Return $x$}
\vspace*{2 mm}
\EndProcedure
\end{algorithmic}
\end{algorithm}


\subsubsection{f-smart method}

We keep track of which variables were flipped in recent improvements. Then, when an escape is needed (in order to diversify the search space), we randomly flip some number of the most-flipped bits. The reasoning behind the flip frequency-based scheme is that certain variables are unstable (defined as being flipped often when updating), and those are precisely the variables we should flip to escape (as opposed to the more-stable variables). We also track the number of escapes with no update to $V_{\tt{min}}$, and increase the number of variables to flip as this number increases, making the flips increasingly drastic when less drastic flips have failed to diversify the search space. This is a version of the adaptive memory idea introduced by Glover et al. \cite{glover1998adaptive, glover1999tabu}.


\subsection{Choice of variables}
\label{VariableChoice}

There are many ways to choose the $k$ variables to solve for from amongst the ${N \choose k}$ possibilities. The simplest one is to choose randomly. However, for all problems we tested there is an advantage to choosing more intelligently. For the benchmarked problems in this paper, a combination of gains-based and fusion-guided (see Sections~\ref{gainsbased} and \ref{fusionguided}) choice of variables was the most advantageous. 


\subsubsection{Gains-based}
\label{gainsbased}

In a 1-opt tabu search, the variable flipped is often the variable with the best one-flip gain. We generalize this such that we try to update the $k$ variables with the best 1-opt gain. This method can get stuck easily if it is not used with tabu search for $k$-opt (see Section~\ref{tabukopt}), since the $k$ variables with the best $k$ one-flip gains will be chosen repeatedly, even when no possible flip can be found that lowers the objective function's value. Another possibility is to choose the $k$ variables randomly, weighting the choice based on the one-flip gains. 


\subsubsection{Fusion-guided}
\label{fusionguided}

This scheme is coupled with path relinking escapes (see Section~\ref{PathRelinking}). Once a pair of parents has been chosen, we can focus our search on the hypercube defined by the symmetric space between these parents. Let $d$ be the Hamming distance between the parents. Then, if $k \leq d$, we choose $k$ variables randomly amongst the $d$ variables; otherwise, we choose all of the $d$ variables and $k-d$ more variables from the set of variables that were equal in the parents. We do this for $w$ iterations, before reverting back to one of the other variable-choice methods: gains, coupling, random, etc. In addition, we note that the tabu search for $k$-opt is not taken into account for these $w$ iterations: the tabu list is kept empty and the candidate list is kept full. 


\subsubsection{Coupling-based}

Intuitively, we might expect it to be advantageous to choose strongly correlated variables to be considered for an update together. There are many ways to do this, and the tradeoff between precision and computational time must be considered. We present a simple example; see Algorithm~\ref{CouplingChoice}. 

This algorithm iterates through two stages until finding $k$ variables out of the candidate variables $C$. In the first stage, a variable is chosen randomly, using probabilities derived from the sum of absolute values of coefficients in that variable's column in the problem matrix $Q$ (we call this the ``strength''). In the second stage, a variable that is connected (in the underlying adjacency matrix) to the variable in the first stage is chosen randomly, using probabilities derived from the absolute value of the coefficients in the first variable's column (we call this the ``conditional strengths''). 

\begin{algorithm}
\caption{Coupling Choice Algorithm}\label{CouplingChoice}
\begin{algorithmic}[1]
\vspace*{2 mm}
\Procedure{CouplingChoice}{}

\State \detokenize{indices_to_update} $\gets \{\}$ (an empty set)

\State $LC \gets$ $C$

\State $LQ \gets$ project $Q$ on $C$

\State $n \gets$ dim($LQ$)

\For {$i=1$ to $n$} strength[$i$] $\gets \sum_j |LQ|_{ji}$ \EndFor

\State probabilities[$i$] $\gets$ strength[$i$] / sum(strength), for each $i$

\While {length(\detokenize{indices_to_update}) < $k$}

    \State $p \gets$ random($0,1$) (first stage)
    
    \State $i \gets C$[index for interval in which $p$ falls in cumulative probabilities]
    
    \If {$i \in LC$} remove $i$ from $LC$ (exclude from second stage) \EndIf
    
    \State Add $i$ to \detokenize{indices_to_update}
    
    \If {length(\detokenize{indices_to_update}) = $k$} break \EndIf
    
    \For {$j \in LC$} \detokenize{conditional_strengths}[$j] \gets |LQ_{ji}|$
    \EndFor
    
    \If {$\sum$(\detokenize{conditional_strengths}) = $0$} continue \EndIf
    
    \State Divide \detokenize{conditional_probabilities} by $\sum$(\detokenize{conditional_strengths})
    
    \State $p \gets$ random($0,1$) (second stage)
    
    \State $m \gets LC$[index for interval in which $p$ falls in cumulative \detokenize{conditional_probabilities}]
    
    \State Remove $m$ from $LC$
    
    \State Add $m$ to \detokenize{indices_to_update}

\EndWhile

\State \Return \detokenize{indices_to_update}

\vspace*{2 mm}

\EndProcedure
\end{algorithmic}
\end{algorithm}


\subsubsection{Practical considerations for the quantum annealer}
\label{DWaveChoice}

The specifics of the graph of the D-Wave chip are also a consideration. For QUBO problems that are more dense than the hardware graph, it is necessary to identify multiple logical qubits with each physical qubit. The mapping between logical and physical qubits is often referred to as an ``embedding'' \cite{choi2008minor, choi2011minor}. 

However, for sparse problems of size $N$, any subproblems of size $k$ are likely to be sparse as well. For this reason, in some scenarios it might be better not to use a complete embedding (forcing $k\leq\sqrt{2n}+1$, where $n$ is the number of qubits), since it might be possible to embed considerably larger subproblems than that ($\sqrt{2n}+1 < k \leq n$). The challenge is minimizing the time expense of finding an embedding for each subproblem. 

Non-complete embeddings can be utilized in at least two ways:
\begin{enumerate}
\item{Find a new embedding before each call. The disadvantage to this method is the time required to find an embedding, and the fact that one cannot know for sure if a graph can be embedded without trying to embed it. It may be possible to devise a heuristic that could judge which size and density combinations are likely to be embeddable, but multiple calls to an embedding finder might still be necessary before an embedding is found and can be used for a call to the quantum annealer. }

\item{Find a pool of subproblems and the corresponding embeddings in advance, and then choose one to use at each step. The advantage to this method is apparent for problems that share the same adjacency matrix, but the coefficients differ. One can then find a pool of subproblems and corresponding embeddings in advance, and at each stage choose one of the prechosen subproblems to optimize over. This method can be combined with the above methods; for example, after determining the $k$ variables with the best one-flip gains, we can choose the subproblem from the pool that contains the largest number of variables from the group of $k$ variables.}

\end{enumerate}


\subsection{Tabu search for $k$-opt}
\label{tabukopt}

We define a version of tabu search for $k$-opt: we maintain a queue of length $TT$ (the tabu tenure), in which each element is a list of variables of length $k$. While the queue is not full we add a new group of $k$ variables at each update. Once the queue is full, we begin to also pop the oldest list of $k$ variables at each step. We also try a version of this in which only the variables that were actually flipped in the update are marked as tabu. In this case the elements in the queue are not of equal length, and indeed are sometimes empty (if no variable out of the $k$ was flipped in the update). 

Unlike 1-opt tabu search, the objective of using the tabu tenure here is not to allow hill climbing, but rather to force the same variables to not be chosen for an update attempt too frequently. The reason for this is that when optimizing over the $k$ variables, at worst the same solution will be chosen for those $k$ variables (based on the way our algorithm is built). Hill climbing occurs from within the underlying optimizer: hills of up to $k$ flips can be climbed, in principle. Since most tabu search algorithms employ a tabu tenure of up to $\sim$20, and typically $k > 20$, we consider this to be sufficient hill climbing capability. 

\subsection{1-opt tabu search improvement}
\label{Tabu1optImprovement}

We allow the option of 1-opt tabu search improvement phases immediately after an escape (before the $k$-opt phase), after convergence (after the $k$-opt phase), or with a probability $p$ of occurring instead of a $k$-opt update (disregarding the tabu queue of the $k$-opt). We typically set the tabu tenure low (or to 0) and the convergence length short, as the idea is to get a quick improvement. We motivate this improvement by noting that despite $k$-opt digging ``deeper'' than 1-opt, the latter has an advantage in that it is able to consider all of the variables at once for an update. When $N$ is much larger than $k$, each variable is visited less frequently by the GQS, and this advantage becomes increasingly important. We note that these optional tabu search 1-opt improvement phases are not indicated in the pseudo-code in the interest of clarity. 


\subsection{Stopping criteria and convergence criterion}
\label{StoppingCriteria}

For benchmarking purposes, we define stopping criteria as a maximum time limit or a maximum number of escapes to be performed. In addition, for some practical problems a known objective function's value is sufficient, so we also allow stopping once a certain value is reached. 

For the convergence criterion, we define convergence as there being no update to the minimum value found since the last escape, within $CL$ (convergence length) steps from the last escape. We also define an objective function's tolerance $tol$. Only prospective updates that lower the objective function's value by more than $tol$ are accepted.


\section{Benchmarking}
\label{benchmarking}


\subsection{Methodology}

For benchmarking purposes, we simulated larger quantum annealers in order to make a claim regarding the scaling of the solution time and/or gap (that is, the difference in the objective function's value from the best known solution). To do this, we replaced D-Wave's machine's being the underlying optimizer with a 1-opt tabu search with short-term memory and aspiration (see, for example, \cite{glover1999tabu}), and then replaced the time the {1-opt} tabu search takes with the expected time it would take the D-Wave solver to solve that problem (we chose 0.02 seconds\footnote{We chose 0.02\,seconds based on 1000 anneals each taking 20\, microseconds (which is the current minimum anneal time). This is the actual time for computation, and does not include extra time for programming the chip, thermalization, etc. Since future run times are not known, this number is only meant as a rough estimate, to give an idea of the actual time the computation could take.}). This allowed us to simulate a D-Wave machine that is many years away (at a significant cost in time). We have verified that the algorithm works with the D-Wave machine as the underlying optimizer, but due to delays as explained in Section~\ref{DifferentMiniSolvers}, there are practical reasons why it is preferable to simulate the quantum annealer for our benchmarking.

The actual time it will take a new D-Wave machine to solve a problem of size $k$ in the future is unknown, and it may be more or less than the number we used. This is due to advances in chip technology, and also depends on the way in which the machine will be used: it is possible to use a small number of reads, a short annealing time, and a short wait before readout to lower the time required. It is also possible that as the number of qubits grows, a longer annealing time may be required in order to maintain the quality of results (this is expected once quantum annealers reach the threshold where if they were to anneal any faster the results would become worse). For these reasons, we also report the number of iterations (calls to the D-Wave machine) which would remain accurate regardless of future run time. The total time for each call is composed of a ``quantum time'', which is the product of the number of iterations and the presumed run time (see above), and a ``classical time'', including all other solver time before and after the underlying optimizer calls. 

For each problem set and each parameter set, we solved each problem 32 times with a time out of 90 seconds of wall-clock time for the problems of size 500, 2250 seconds for the problems of size 2500, and $1.2N$ seconds for the problems of size 3000 and above. We report the time to best solution, the average relative gap between the best solution found compared to the best known solution, the number of underlying optimizer iterations to the best solution, and the success rate (the percentage of repetitions that reached the best known solution). 

We benchmarked the GQS on three sets of problems. The first is the OR-Library unconstrained binary quadratic programming problems \cite{orlib}. We generated these problems (labelled ``bqp'' below) by randomly choosing integers between $-100$ and $+100$ from a uniform distribution such that the resulting problem has a density of 0.1 (that is, only 10\% of the elements are non-zero). They have been widely benchmarked in the combinatorial optimization literature. We used the best known solutions from Tavares \cite{tavares2008new}. 

The second set of problems, random fully dense QUBO problems (labelled ``RFDQ'' below) was generated in the same way, but they have a density of 1. We obtained the best known solutions by running a multi-start 1-opt tabu search solver with a large number of iterations, as well as performing a second check by running the path relinking algorithm of Wang et al. \cite{wang2012path}. The third set of problems was generated using the problem generator and settings written by Palubeckis\footnote{ The source code of the generator and the input files to create these problems can be found at \\
\url{http://www.proin.ktu.lt/~gintaras/ubqop_its.html}. This page was last retrieved on July 21, 2015.} \cite{palubeckis2004multistart} with densities ranging from 0.5 to 1, and we used the best known solutions reported in Wang et al.\cite{wang2012path}.

All benchmarking was performed on Amazon Web Services using a c3.8xlarge instance, which has 32 virtual CPUs running at 2800\,MHz. Each call to the GQS was run serially on a single CPU. The code was written in Python, utilizing optimized external libraries such as Numpy and Scipy. An efficient implementation written in a lower-level language such as C would give faster times. However, we expect our results to still hold qualitatively, and much of the prospective time advantage of a lower-level programming language is cancelled by our method of replacing the true run time of the underlying optimizer with the assumed run
time of the D-Wave annealer (the number of iterations would remain the same). 


\subsection{Parameters}

In the results below, we set the parameters that were not varied as listed in Table~\ref{BenchmarkingParametersConstant}. The settings were chosen empirically from problems of sizes up to 500 from OR-Library as well as prior literature (such as \cite{wang2012path}). The names and descriptions of the parameters that were varied are in Table~\ref{BenchmarkingParametersVarying}.

\begin{table}[htdp]
\centering
\caption{Benchmarking parameters (constant)}
\vspace*{2 mm}
\begin{tabular}{|l|c|l|}
\hline
Name & Value & Description \\
\hline
\detokenize{optimizer_TT} & 15--20 & tabu tenure for the 1-opt tabu search underlying optimizer \\
\detokenize{optimizer_CL} & 10$k$ & \pbox{20cm}{maximum iterations with no improvement for the 1-opt tabu search \\ underlying optimizer} \\
\detokenize{parent_distance_threshold} & 5 & minimum Hamming parent-parent distance \\
\detokenize{child_distance_threshold} & 0.33 & minimum Hamming parent-child distance \\
\detokenize{num_elite_solutions} & 10 & number of elite solutions to keep \\
\detokenize{variable_choice} & ``gains'' & variable choice type after $w$ iterations (see below) \\
\detokenize{presumed_time} & 0.02\,sec & presumed run time for quantum annealer \\
\detokenize{tol} & $10^{-8}$ & minimum change in value for an update \\
\hline
\end{tabular}
\label{BenchmarkingParametersConstant}

\end{table}

\begin{table}[htdp]
\centering
\caption{Benchmarking parameters (varying)}
\vspace*{2 mm}
\begin{tabular}{|l|l|}
\hline
Name & Description \\
\hline
$\detokenize{k}$ & variable group size for underlying optimizer \\
$\detokenize{CL}$ & maximum calls to the underlying optimizer with no improvement \\
$\detokenize{w}$ & number of fusion-guided iterations \\
$\detokenize{TT}$ & $k$-opt tabu tenure, typically $\propto N/k$ \\
$\detokenize{W}$ & tabu whole group (True/False) \\
\hline
\end{tabular}
\label{BenchmarkingParametersVarying}
\end{table}

In the case of a quantum annealer, $k$ refers to the largest complete problem that can be solved with that quantum annealer. Depending on the physical connectivity of the chip, the number of qubits in the quantum annealer $n$ could be much larger (see Section~\ref{QuantumAnnealer}).


\subsection{Results}

\subsubsection{OR-Library and RFDQ: 500 to 2500 variables}

We present results for the OR-Library 500 and 2500 problems and RFDQ problems of size 500 below. We benchmarked the smaller instances comprehensively to choose the best settings, optimizing for highest success rate, which we then used for all remaining benchmarks. These runs were based on the path relinking escapes, with an initial reference set initialized from random, and updated using an underlying optimizer of size 50 with a convergence length of 3 and a $k$-opt tabu tenure of 6 (leading us to use $TT=0.6N/k$ hereafter), marking as tabu all variables chosen at each update ($W=$ True). The variables chosen to be updated were based on the fusion-guided approach with $w=1$, and based on the best one-flip gains thereafter (and during the creation of the reference set). The underlying optimizer was 1-opt tabu search with a tabu tenure of 15 and convergence length 500. We verified that the algorithm is not very sensitive to the exact parameter values chosen: the differences are small, and are close to the statistical variation of the results. 

Table~\ref{TableSummarizedResults500} includes the results for each problem set, averaged over the 10 problems in each set, as well as 32 repetitions for each problem. Detailed results for each problem (averaged over the 32 repetitions) in the OR-Library and RFDQ problem sets are presented in Table~\ref{TableDetailedResultsORLib500} and Table~\ref{TableDetailedResultsRFDQ500}, respectively.

\begin{table}[htdp]
\centering
\caption{Results for size $500$ for $k=50$, $CL=3$, $TT=64$, $w=1$, fusion-guided path relinking with $W=$ True. The average time to best solution in seconds (assuming an underlying optimizer call time of 0.02\,seconds) is denoted <$T$>, the average gap to the best known solution as a percentage is denoted <$G$>, the fraction of GQS calls that resulted in finding the best known solution is denoted ``succ.'', and the average number of calls to the underlying optimizer is denoted <$I$>.}
\begin{tabular}{|l|r|r|r|r|r|r|r|}
\hline
problem set &   <$T$>   & STD($T$)  &   <$G$>   & STD($G$)  &  succ.  & <$I$>  & STD($I$) \\
\hline
bqp500   &    4.83 &    2.60 &    0.02 &    0.06 &  60.62 &  158.3 &  91.2  \\
RFDQ500   &    2.98 &    2.50 &    0.01 &    0.04 &  78.75 &   95.9 &  87.8 \\
\hline
\end{tabular}
\label{TableSummarizedResults500}
\end{table}

\begin{table}[htdp]
\centering
\caption{Detailed results for OR-Library $500$ for $k=50$, $CL=3$, $TT=6$, $w=1$, fusion-guided path relinking with $W=$ True. See the caption of Table~\ref{TableSummarizedResults500} for a definition of <$T$>, <$G$>, ``succ.'', and <$I$>.}
\begin{tabular}{|l|r|r|r|r|r|r|r|}
\hline
name       &   <$T$>   & STD($T$)  &  <$G$>   & STD($G$) & succ.  &  <$I$>   & STD($I$) \\ 
\hline
bqp500-1   &    5.85 &    2.51 &   0.10 &   0.13 &  50.00 &  197.3 &   87.9 \\
bqp500-2   &    3.74 &    2.54 &   0.00 &   0.01 &  90.62 &  118.3 &   88.3 \\
bqp500-3   &    1.55 &    1.00 &   0.00 &   0.00 & 100.00 &   45.8 &   33.7 \\
bqp500-4   &    5.93 &    1.89 &   0.01 &   0.01 &  34.38 &  188.2 &   67.1 \\
bqp500-5   &    3.52 &    1.82 &   0.00 &   0.02 &  96.88 &  113.7 &   64.1 \\
bqp500-6   &    6.25 &    1.80 &   0.04 &   0.04 &  46.88 &  206.3 &   64.2 \\
bqp500-7   &    6.35 &    2.10 &   0.02 &   0.02 &  37.50 &  209.4 &   75.6 \\
bqp500-8   &    6.34 &    2.10 &   0.04 &   0.02 &  12.50 &  211.5 &   76.7 \\
bqp500-9   &    5.96 &    1.99 &   0.02 &   0.04 &  40.62 &  204.9 &   71.1 \\
bqp500-10  &    2.85 &    1.75 &   0.01 &   0.06 &  96.88 &   87.2 &   58.7 \\
\hline
\end{tabular}
\label{TableDetailedResultsORLib500}
\end{table}

\begin{table}[htdp]
\centering
\caption{Detailed results for RFDQ $500$ for $k=50$, $CL=3$, $TT=8$, $w=1$, fusion-guided path relinking with $W=$ True. See the caption of Table~\ref{TableSummarizedResults500} for a definition of <$T$>, <$G$>, ``succ.'', and <$I$>.}
\begin{tabular}{|l|r|r|r|r|r|r|r|}
\hline
name       &   <$T$>   & STD($T$)  &  <$G$>   & STD($G$) & succ.  &  <$I$>   & STD($I$) \\ 
\hline
RFDQ500-1  &    3.91 &    2.59 &   0.08 &   0.03 &  12.50 &  129.7 &   91.1 \\
RFDQ500-2  &    4.71 &    2.94 &   0.01 &   0.01 &  62.50 &  156.8 &  104.5 \\
RFDQ500-3  &    1.25 &    0.77 &   0.00 &   0.00 & 100.00 &   36.2 &   24.1 \\
RFDQ500-4  &    3.74 &    1.95 &   0.03 &   0.06 &  75.00 &  121.7 &   67.2 \\
RFDQ500-5  &    1.52 &    1.01 &   0.00 &   0.00 & 100.00 &   44.7 &   33.1 \\
RFDQ500-6  &    5.03 &    2.69 &   0.02 &   0.05 &  84.38 &  173.2 &   95.3 \\
RFDQ500-7  &    3.07 &    2.62 &   0.01 &   0.01 &  68.75 &   95.5 &   89.1 \\
RFDQ500-8  &    1.26 &    0.76 &   0.00 &   0.00 & 100.00 &   34.6 &   23.2 \\
RFDQ500-9  &    3.94 &    2.43 &   0.01 &   0.01 &  84.38 &  128.7 &   86.2 \\
RFDQ500-10 &    1.32 &    0.86 &   0.00 &   0.00 & 100.00 &   37.8 &   28.2 \\
\hline
\end{tabular}
\label{TableDetailedResultsRFDQ500}
\end{table}

\begin{table}[htdp]
\centering
\caption{Results for size $2500$ for $k=50$, $CL=3$, $TT=30$, $w=1$, fusion-guided path relinking with $W=$ True. See the caption of Table~\ref{TableSummarizedResults500} for a definition of <$T$>, <$G$>, ``succ.'', and <$I$>.}
\begin{tabular}{|l|r|r|r|r|r|r|r|}
\hline
identifier       &   <$T$>   & STD($T$)  &   <$G$>   & STD($G$)  & succ.  &  <$I$>   & STD($I$) \\ 
\hline
bqp2500  &  173.07 &   90.76 &    0.02 &    0.02 &  34.06 & 3798.0 & 2008.6 \\
\hline
\end{tabular}
\label{TableSummarizedResults2500}
\end{table}

In addition, Table~\ref{TableSummarizedResults2500} includes the results for OR-Library 2500, averaged over the 10 problems in each set, as well as 32 repetitions for each problem. Detailed results for each problem (averaged over the 32 repetitions) in the OR-Library problem sets are presented in Table~\ref{TableDetailedResultsORLib2500}.

\begin{table}[htdp]
\centering
\caption{Detailed results for RFDQ $500$ for $k=50$, $CL=3$, $TT=30$, $w=1$, fusion-guided path relinking with $W=$ True. See the caption of Table~\ref{TableSummarizedResults500} for a definition of <$T$>, <$G$>, ``succ.'', and <$I$>.}
\begin{tabular}{|l|r|r|r|r|r|r|r|}
\hline
name       &   <$T$>   & STD($T$)  &  <$G$>   & STD($G$) & succ.  &  <$I$>   & STD($I$) \\ 
\hline
bqp2500-1  &  180.63 &   82.01 &   0.01 &   0.01 &  34.38 & 3966.6 & 1804.7 \\
bqp2500-2  &  197.50 &   95.32 &   0.05 &   0.02 &   0.00 & 4381.8 & 2091.1 \\
bqp2500-3  &  190.15 &   74.70 &   0.03 &   0.02 &   9.38 & 4240.1 & 1679.0 \\
bqp2500-4  &   77.32 &   49.22 &   0.00 &   0.00 & 100.00 & 1634.7 & 1070.9 \\
bqp2500-5  &  140.69 &   87.81 &   0.00 &   0.00 &  62.50 & 3033.3 & 1915.4 \\
bqp2500-6  &  198.85 &   80.40 &   0.01 &   0.01 &  25.00 & 4334.7 & 1761.6 \\
bqp2500-7  &  245.32 &   67.62 &   0.03 &   0.02 &  18.75 & 5387.8 & 1472.5 \\
bqp2500-8  &  170.00 &   85.64 &   0.01 &   0.01 &  46.88 & 3716.1 & 1866.9 \\
bqp2500-9  &  148.07 &   83.10 &   0.01 &   0.01 &  43.75 & 3241.9 & 1854.1 \\
bqp2500-10 &  182.20 &   88.32 &   0.05 &   0.02 &   0.00 & 4043.4 & 1965.8 \\
\hline
\end{tabular}
\label{TableDetailedResultsORLib2500}
\end{table}


\subsubsection{Palubeckis: 3000 to 7000 variables}

Results for the Palubeckis instances of sizes 3000--7000 are presented in Table~\ref{TableDetailedResultsPalubeckis}. 

\begin{table}[htdp]
\centering
\caption{Detailed results for Palubeckis instances of size  3000--7000 for $k=50$, $CL=3$, $w=1$, fusion-guided path relinking with $W=$ True. See the caption of Table~\ref{TableSummarizedResults500} for a definition of <$T$>, <$G$>, ``succ.'', and <$I$>.}
\begin{tabular}{|l|r|r|r|r|r|r|r|}
\hline
name       &   <$T$>   & STD($T$)  &  <$G$>   & STD($G$) & succ.  &  <$I$>   & STD($I$) \\
\hline
p3000-1    &  292.89 &  135.37 &   0.03 &   0.03 &  40.62 & 5967.0 & 2770.8 \\
p3000-2    &  280.17 &  124.55 &   0.01 &   0.01 &  18.75 & 5614.8 & 2486.6 \\
p3000-3    &  313.99 &  153.08 &   0.03 &   0.02 &  15.62 & 6367.3 & 3080.9 \\
p3000-4    &  368.17 &  148.97 &   0.04 &   0.01 &   3.12 & 7423.2 & 2978.1 \\
p3000-5    &  323.24 &  142.11 &   0.03 &   0.01 &   0.00 & 6493.5 & 2843.6 \\
\hline
p3000  &  315.69 &  144.39 &    0.03 &    0.02 &  15.62 & 6373.2 & 2903.9  \\
\hline\hline
p4000-1    &  599.18 &  143.46 &   0.02 &   0.03 &  43.75 & 10431.8 & 2481.5 \\
p4000-2    &  466.75 &  187.62 &   0.05 &   0.02 &   6.25 & 8018.7 & 3202.9 \\
p4000-3    &  544.72 &  151.48 &   0.04 &   0.02 &   0.00 & 9430.6 & 2619.2 \\
p4000-4    &  403.19 &  210.82 &   0.02 &   0.02 &  40.62 & 6996.3 & 3650.3 \\
p4000-5    &  564.33 &  184.65 &   0.06 &   0.02 &   0.00 & 9680.3 & 3147.6 \\
\hline
p4000  &  515.63 &  191.05 &    0.04 &    0.03 &  18.12 & 8911.5 & 3290.8  \\
\hline \hline
p5000-1    &  773.82 &  276.58 &   0.06 &   0.03 &   0.00 & 11708.2 & 4163.0 \\
p5000-2    &  614.80 &  263.88 &   0.04 &   0.02 &   0.00 & 9254.3 & 3917.7 \\
p5000-3    &  623.26 &  251.04 &   0.07 &   0.03 &   3.12 & 9459.2 & 3739.8 \\
p5000-4    &  783.66 &  228.80 &   0.07 &   0.03 &   0.00 & 11714.3 & 3402.6 \\
p5000-5    &  774.58 &  211.93 &   0.05 &   0.02 &   0.00 & 11426.0 & 3088.7 \\
\hline
p5000  &  714.02 &  259.46 &    0.06 &    0.03 &   0.62 & 10712.4 & 3846.7  \\
\hline \hline 
p6000-1    & 1018.34 &  302.98 &   0.08 &   0.04 &   0.00 & 13584.2 & 3974.2 \\
p6000-2    &  944.05 &  246.74 &   0.06 &   0.03 &   0.00 & 12556.8 & 3255.0 \\
p6000-3    &  955.48 &  293.04 &   0.06 &   0.03 &   0.00 & 12813.8 & 3909.6 \\
\hline
p6000  &  972.62 &  283.87 &    0.07 &    0.03 &   0.00 & 12985.0 & 3752.6  \\
\hline \hline
p7000-1    & 1215.57 &  362.55 &   0.06 &   0.02 &   0.00 & 14539.8 & 4285.1 \\
p7000-2    & 1209.88 &  371.84 &   0.07 &   0.02 &   0.00 & 14574.3 & 4373.1 \\
p7000-3    & 1179.84 &  337.17 &   0.09 &   0.02 &   0.00 & 14225.0 & 3940.6 \\ 
\hline
p7000  & 1201.76 &  357.83 &    0.07 &    0.02 &   0.00 & 14446.4 & 4206.7  \\
\hline 
\end{tabular}
\label{TableDetailedResultsPalubeckis}
\end{table}


\subsubsection{Summary of all results: 500 to 7000 variables}

\begin{figure}[!htb]
\centering
\begin{overpic}[width=0.49\textwidth]{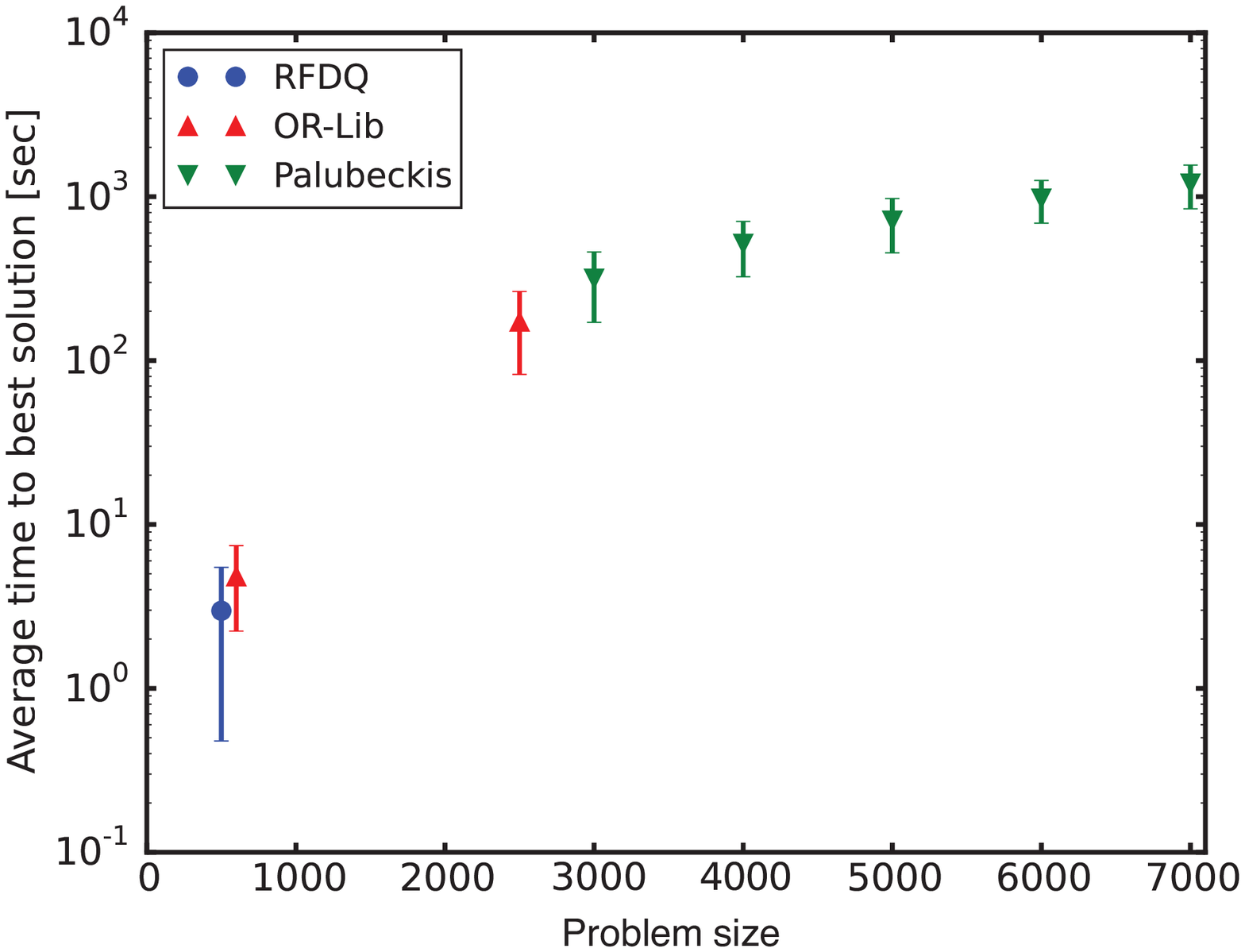}
 \put (0,69) {$(a)$}
\end{overpic}
\begin{overpic}[width=0.49\textwidth]{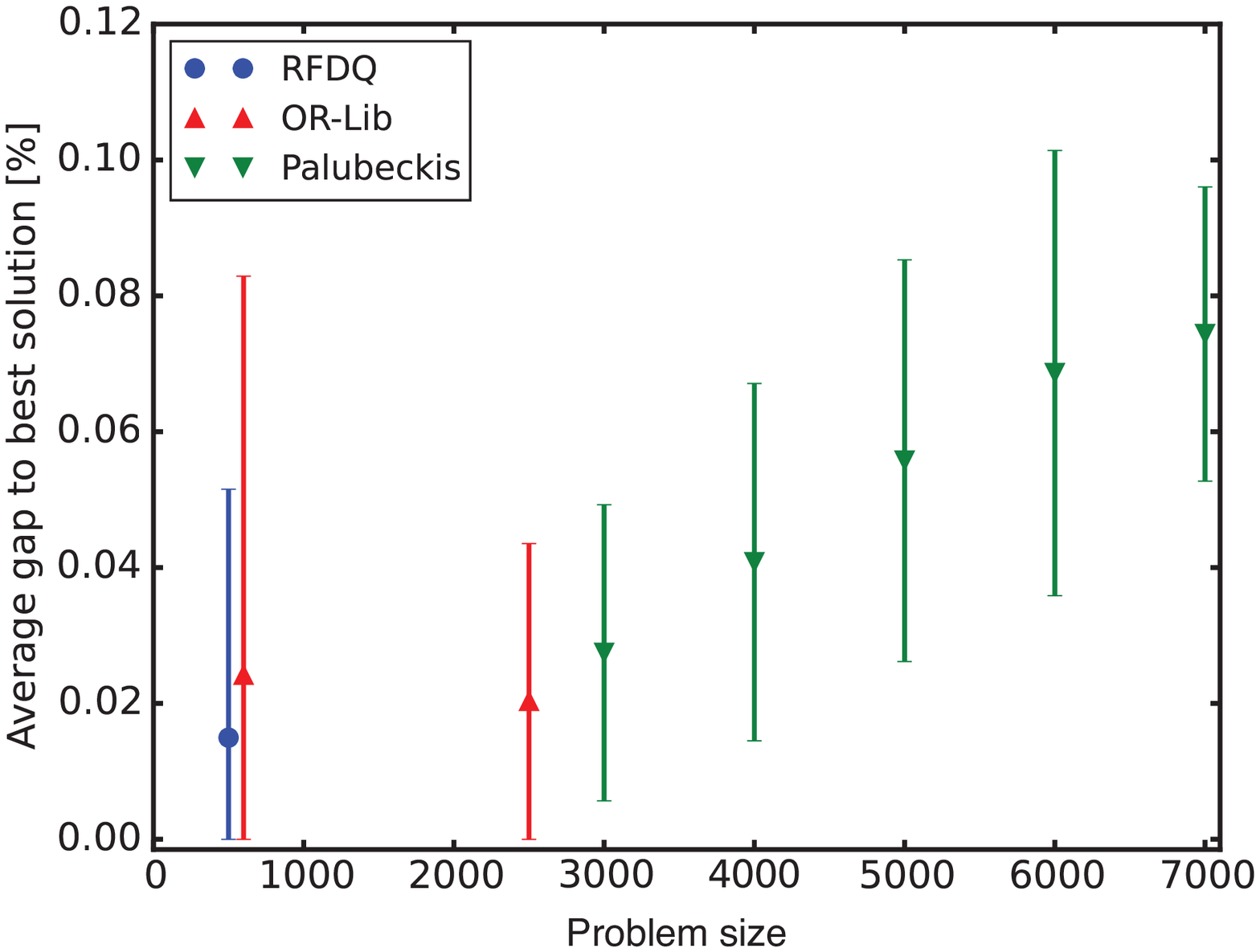}
 \put (0,69) {$(b)$}
\end{overpic}
\caption{(a) Average time as a function of the problem size. Data was collected for GQS with parameters $k=50$, $CL=3$, $w=1$, $TT = 0.6N/k$, fusion-guided path relinking with $W=$ True, averaged over the respective problems and 32 repetitions per problem. (b) Average gap as a function of problem size. Note for both figures: the results for OR-Library 500 were moved slightly to the right for clarity.}
\label{fig:vs_N}
\end{figure}

We summarize all of the results for average time and average gap in Fig.~\ref{fig:vs_N}.

\subsubsection{Benchmarking for practical requirements}

For practical use, we expect that for many use cases, once a solution has been found that is close enough to the global optimum, one can stop the search based on stopping criteria, even if the confidence in having successfully found the global optimum is not high. The user can define how close is sufficient. With this in mind, we can run the solver such that it stops once we find a solution with a given gap to the best known solution. In practice, the average time (and distribution of times) necessary to do this for a given set of problems could then be used as a guideline for the time it would require to attain that level of quality of solutions to new problems (assuming they are of similar difficulty). 

As an example, we consider that for many applications, once a local optimum has been found with a gap of, for example, $0.1\%$, there is little benefit in searching any further. One reason for this is that the parameters of the problem often have considerable uncertainties associated with them, leading to an uncertainty in the objective function's value that is larger than the gap itself. In other applications, such as machine learning, it is undesirable to ``over-fit'', by finding the very best fit of the model to the training data, because it will not generalize to the next set of data, and for this reason optimizers will often intentionally be stopped early.

\begin{figure}[!htb]
\centering
\begin{overpic}[width=0.49\textwidth]{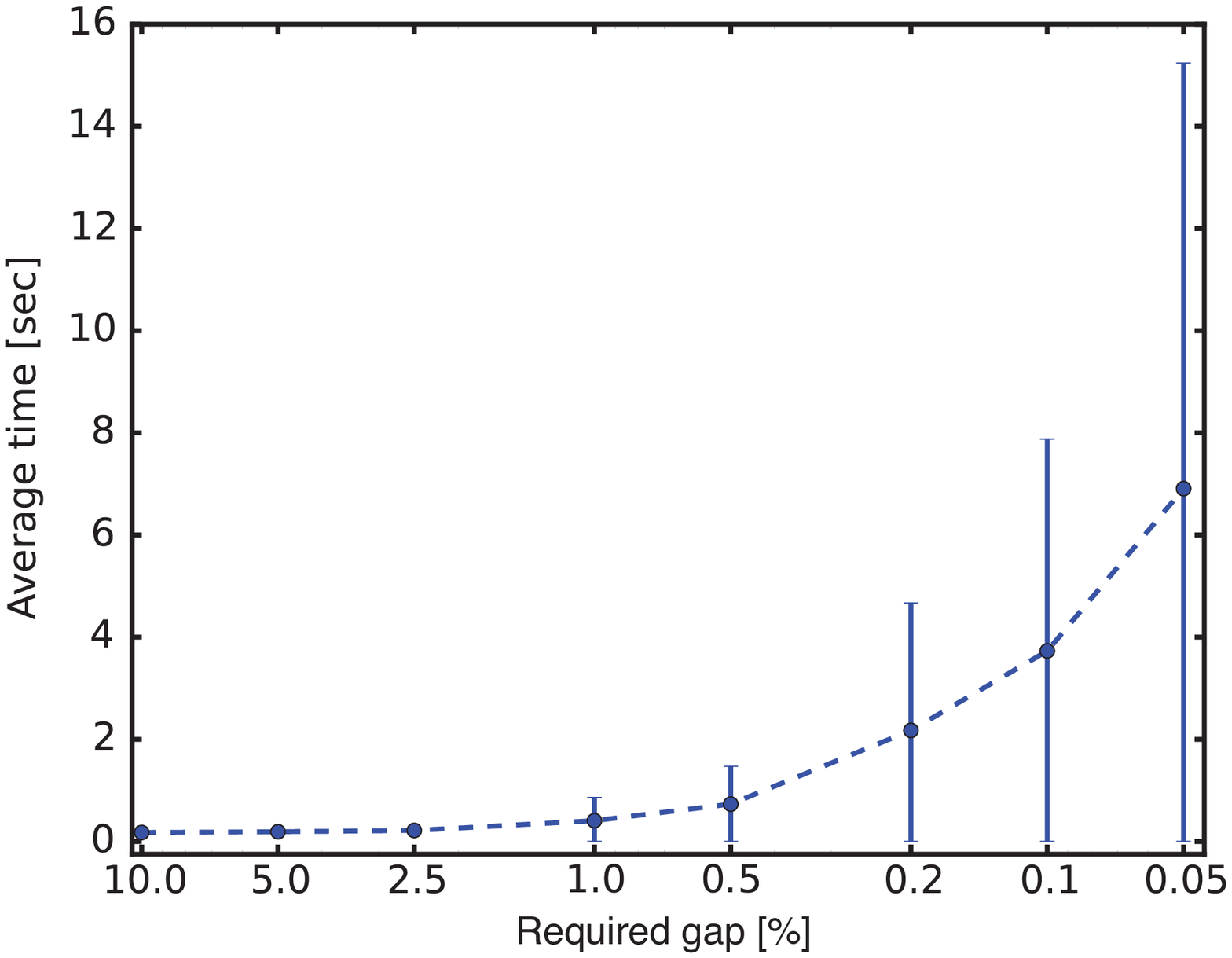}
 \put (0,69) {$(a)$}
\end{overpic}
\begin{overpic}[width=0.49\textwidth]{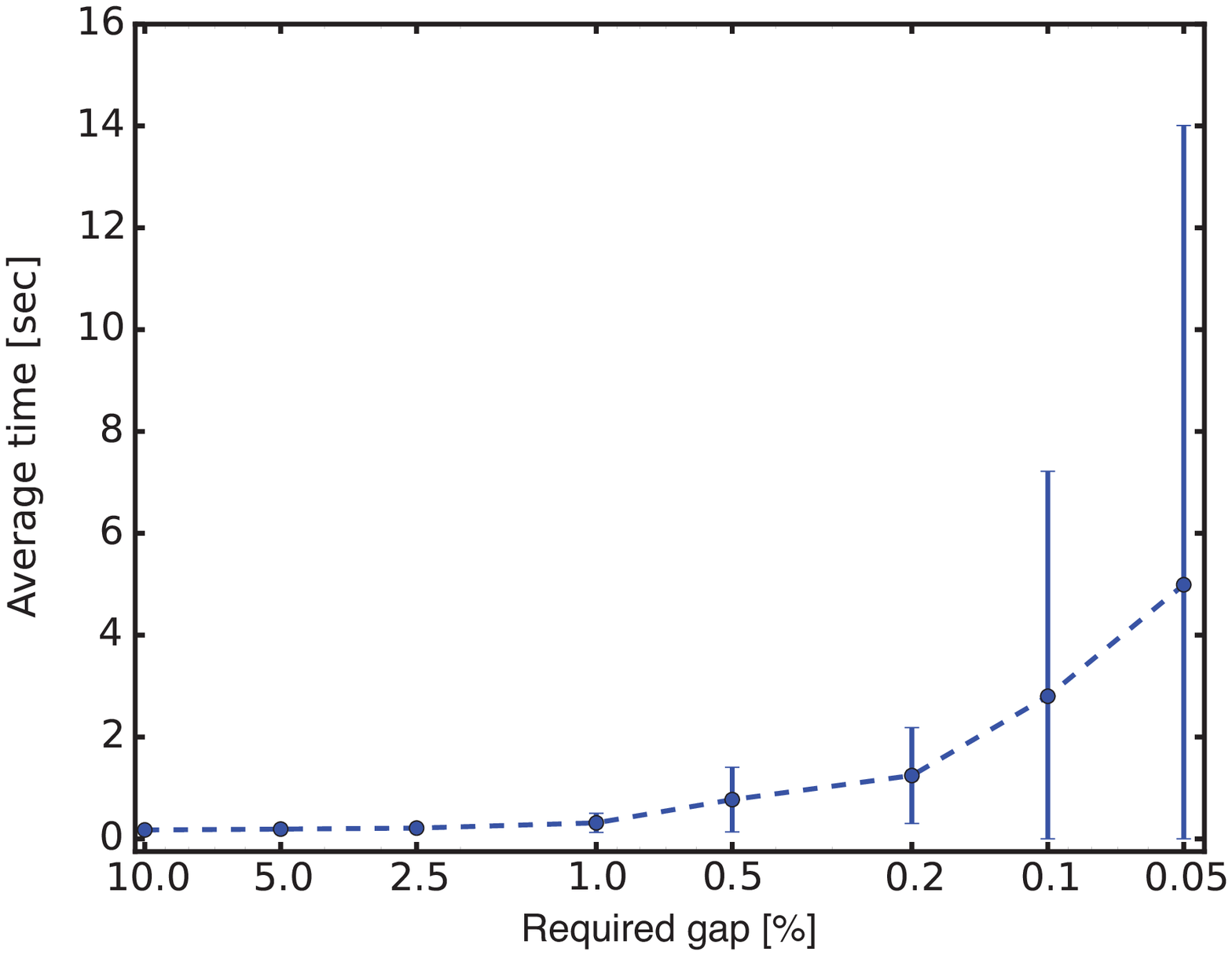}
 \put (0,69) {$(b)$}
\end{overpic}
\caption{(a) Average time as a function of the required gap. Data was collected for GQS with parameters $k=50$, $CL=3$, $TT=8$, $w=1$, fusion-guided path relinking with $W=$ True, for OR-Library problems of size 500, averaged over the ten problems and 32 repetitions. (b) The same for RFDQ problems of size 500.}
\label{fig:time_vs_gap}
\end{figure}

We investigated how the time required varied as we change the desired gap. Fig.~\ref{fig:time_vs_gap} provides results for the OR-Library and RFDQ problems of size 500. For these problems, we found that the time required to find a solution that is, for instance, $1\%$ from the best known solution is low. In addition, as we require finding a solution closer to the best known solution, the time rises rapidly, as does the variation in the time required to find said solution, for both types of problems. 
 
\subsubsection{Dependence on size of underlying optimizer}
\label{DependenceMiniSolverSize}

\begin{figure}[!htb]
\centering
\begin{overpic}[width=0.49\textwidth]{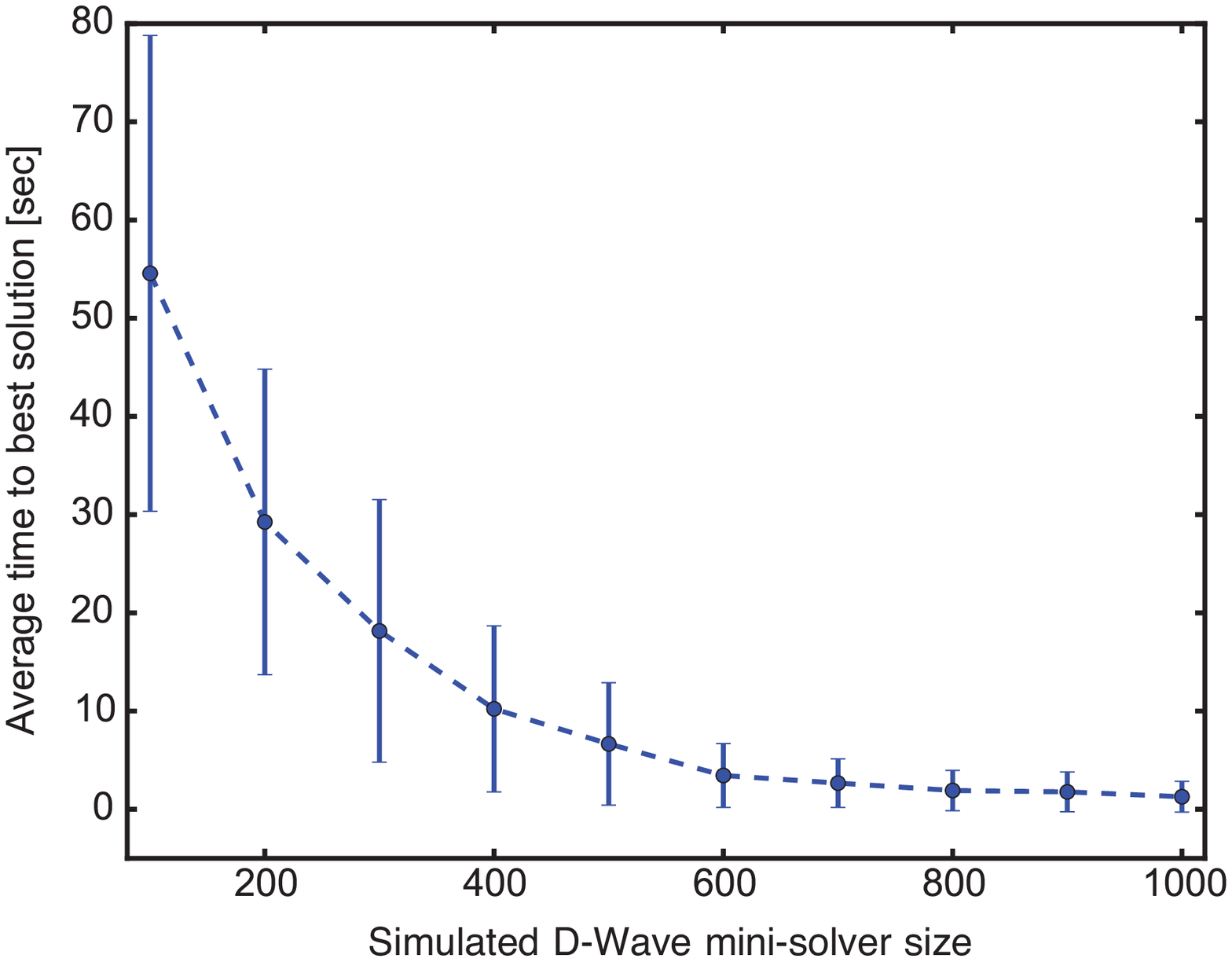}
 \put (0,69) {$(a)$}
\end{overpic}
\begin{overpic}[width=0.49\textwidth]{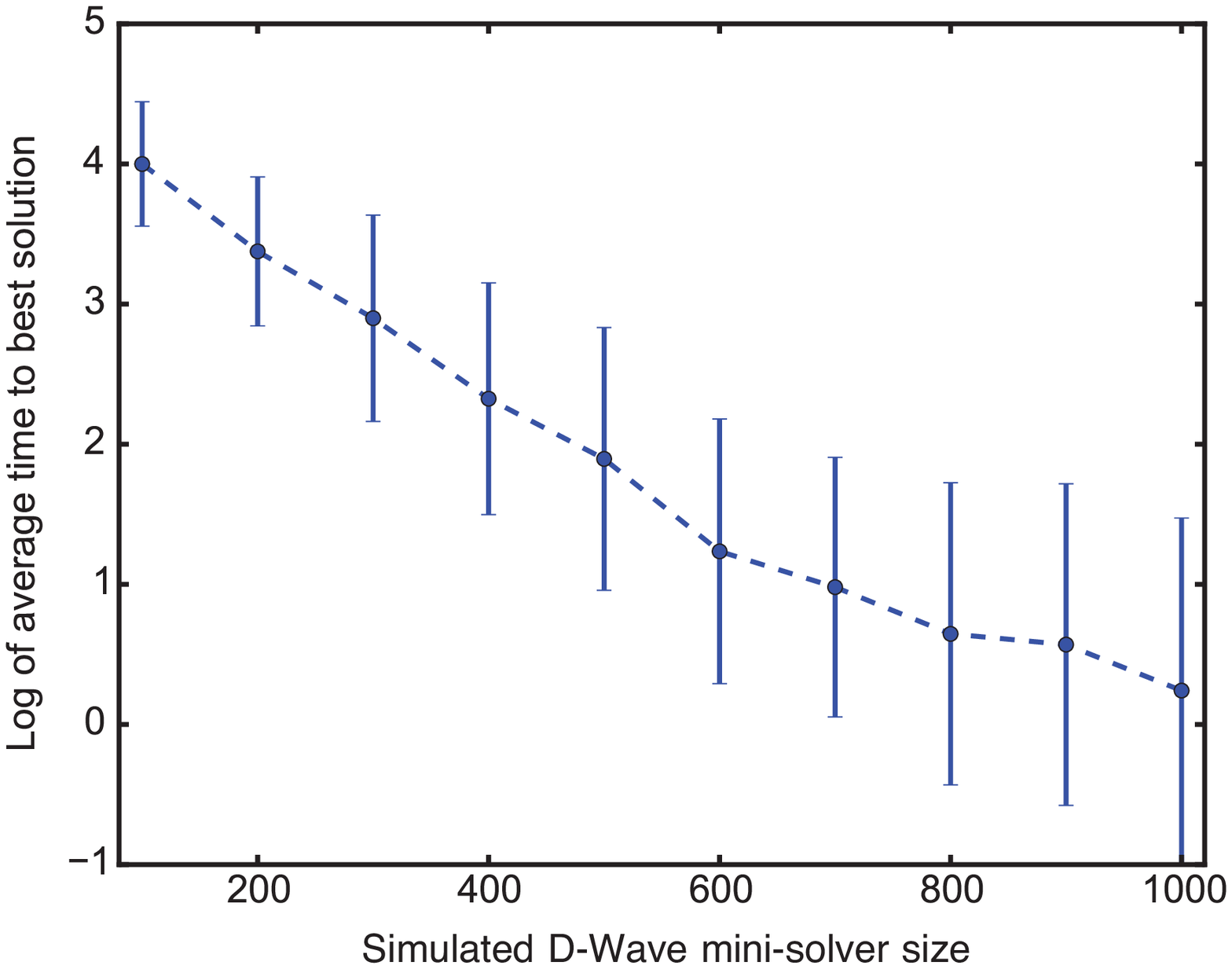}
 \put (-2,69) {$(b)$}
\end{overpic}
\caption{(a) Average time to best solution as a function of $k$, the simulated D-Wave chip size. Data was collected for GQS with parameters $CL=3$, $w=1$, fusion-guided path relinking with $W=$ True for OR-Library problems of size 2500, averaged over the ten problems and 32 repetitions, and $TT$ was chosen such that the tabu list is $60\%$ of the problem size. (b) Log of the average time to best solution as a function of the underlying optimizer size (same data as (a)). }
\label{fig:time_vs_k}
\end{figure}

We investigated the dependence of the average time to best solution on the underlying optimizer size; see Fig.~\ref{fig:time_vs_k}(a). We found that the time to best solution falls exponentially, as can be seen by observing that the time to best solution falls linearly as a function of the logarithm of the underlying optimizer size; see Fig.~\ref{fig:time_vs_k}(b). The expectation is that as $k \to N$, the number of iterations required should go to 1, giving a time equal to the presumed time for one call to the underlying optimizer (in this case 0.02 seconds). The exponential decrease can be understood by observing that as the underlying optimizer increases in size, it is able to scan an exponentially increasing variable space in constant time (based on our assumption of constant time). 

\begin{figure}[!htb]
\centering
\includegraphics[scale=.4]{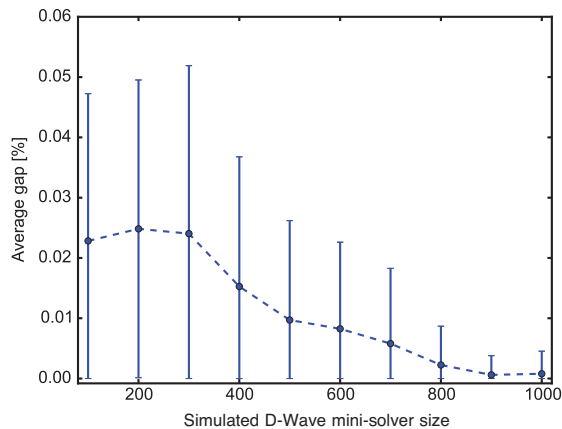}
\caption{Average gap as a function of the simulated D-Wave chip size, with the same parameters as in Fig.~\ref{fig:time_vs_k}.}
\label{fig:gap_vs_k}
\end{figure}

We also investigated the dependence of the average gap obtained on the underlying optimizer size; see Fig.~\ref{fig:gap_vs_k}. We see that the gap decreases slowly. Intuitively, as the underlying optimizer size is increased, the GQS considers flipping larger and larger subgroups of bits, which contain within them the bit flips for any smaller underlying optimizers. Hence, we expect the gap to stay constant at worst, and at best to improve.

\section{Discussion}
\label{discussion}

We have also tried solving the two sets of problems bqp500 and RFDQ500 with a 1-opt tabu search with a tabu tenure of 20 and a convergence length of 2500 with 100 repetitions per problem (see Table~\ref{TableTabuResults}). The results show that the time to best solution is only slightly higher for the dense problems (recall that the OR-Library problems have a density of 0.1), but the success rate is much higher and the gap is lower. Based on this, finding a good local optimum appears to be similarly hard in both cases (for tabu 1-opt search), but finding the global optimum is harder in the sparse problems. For most practical applications, we expect the difference between the gaps ($0.02\%$ and $<0.005\%$) to be negligible, so for practical purposes it appears that these two sets of problems are of similar difficulty. 

\begin{table}[htdp]
\centering
\caption{1-opt tabu results. See the caption of Table~\ref{TableSummarizedResults500} for a definition of <$T$>, <$G$>, and ``succ''. The average number of one-bit flips is denoted <I>.}
\begin{tabular}{|c|c|c|c|c|c|c|c|}
\hline
set       &   <$T$>   & STD($T$)  &  <$G$>   & STD($G$) & succ.  & <$I$>  & STD($I$) \\
\hline 
bqp500   &   0.30  &   0.19 &   0.02 &       0.02 & 52.00 &  1347 &   974 \\
RFDQ500    &    0.38 &   0.30 &   0.00 &   0.01 & 95.00 &   1837 &    1596 \\
\hline
\end{tabular}
\label{TableTabuResults}
\end{table}

We note that one might have expected that the dense problems would be harder, but they do not appear to be. To explain this, we hypothesize that sparse random problems contain many more local minima (relative to the one-flip neighbourhood), and hence the optimizer can more easily get stuck in a local minimum and miss the global optimum. A possible explanation for this is that given a solution and its objective function's value, in a sparse problem the probability of flipping some number of bits and ending up at a similar objective function's value is much higher than in a dense problem. It follows that the probability of the existence of multiple local minima with a similar value of the objective function is much greater. 

Comparing the results of GQS in Table~\ref{TableSummarizedResults500} (where $k=50$) with the results for the 1-opt tabu search in Table~\ref{TableTabuResults}, we note first that the GQS results are marginally better for the OR-Library problems (that is, they have a higher success rate) and marginally worse for the RFDQ problems. Secondly, we note that the time to best solution is considerably longer for the GQS for both sets of problems.

As we increase the underlying optimizer's size, we expect the average time to best solution to drop and the quality of the solution (that is, the average gap) to either decrease or remain constant. Based on this expectation, we predict that when the underlying optimizer's (the D-Wave chip) size is large enough, it will eventually beat 1-opt tabu search and all other classical solvers (see Section~\ref{DependenceMiniSolverSize}). Exactly when this occurs will depend on the quality and run time of the D-Wave processor. 


\section{Conclusions and Future Work}
\label{conclusions}

We have shown that it is possible to use the D-Wave machine to heuristically solve significantly larger problems than the chip intrinsically allows. If in the future the D-Wave machine is able to optimize QUBO problems faster than the best classical algorithms, this contribution could make the machine more suitable for solving real-world problems earlier.
Based on our approach, we found that sparse and dense random problems of size 500 can be optimized using a $k=50$ simulated quantum annealer to within $0.02\%$ of the optimum in 100--160 iterations, assuming the quantum annealer's solution quality is similar to 1-opt tabu search. If we assume that the run time is $0.02$\,seconds, solving these problems requires $3$--$5$ seconds. Keeping the solver size constant, for larger problem sizes, the number of iterations increases rapidly and the quality of the produced solutions degrades, although it remains within $0.07\%$ for all problems, even for the dense problems of size 7000 (and would likely reduce further if we increased the time out). We would expect these results to improve when using a better underlying optimizer (that is, one with a higher success rate). 

Although the time to best solution for a D-Wave chip with a complete graph of size $k=50$ is not competitive, we have shown that as $k$ grows, the time required to solve a given problem drops exponentially, assuming constant call time to quantum annealer. 
Based on this assumption, and assuming quantum annealers scale faster than classical algorithms on classical hardware, we would expect that the D-Wave chip will eventually beat 1-opt tabu search and all other classical solvers. Exactly when this will occur will depend on the quality and run time of the processor. 

Our estimate of the time that will be required in the future for a D-Wave machine to solve different problems has an unknowable precision since it depends on future advances in engineering which could change the run time of the machine. Similarly, future advances in error correction \cite{pudenz2014error, pudenz2014quantum, kelly2014state} and expected decreases in intrinsic error will affect the quality of the quantum annealer's results. To the extent that this quality is better than the underlying optimizer we used to simulate the D-Wave machine, the actual D-Wave machine's solving time could decrease faster than estimated. 

One way in which our algorithm could be improved is by implementing preprocessing. There are many ways in which QUBO problems can be preprocessed to decrease the size of the problem (several are reviewed in \cite{tavares2008new}). In addition, using a better underlying optimizer such as the path relinking algorithm would undoubtedly improve the results \cite{wang2012path}. Finally, we look forward to improvements in the D-Wave machine in the coming years, at which point we hope to see competitive results that use the machine itself as the underlying optimizer (and possibly using the embedding ideas from Section~\ref{DWaveChoice}). \\


\small \noindent
\textbf{Acknowledgments}    
\hspace{0.25cm} The authors would like to thank Marko Bucyk for editing a draft of this paper and Robyn Foerster, Phil Goddard, and Pooya Ronagh for their useful comments. This work was supported by 1QB Information Technologies (1QBit) and Mitacs.

\small \noindent
\textbf{Conflict of Interest}    
\hspace{0.25cm} EH declares no conflict of interest. GR and BW are academic interns at 1QBit and MV was an academic intern at 1QBit when the work was done. 1QBit is focused on solving real world problems using quantum computers. D-Wave Systems is a minority investor in 1QBit.  


\bibliography{GeneralSolverFull}{}
\bibliographystyle{ieeetr} %

\end{document}